\documentclass[
 amsmath,amssymb,aps,prb,twocolumn
]{revtex4}

\usepackage{bm}
\usepackage[pdftex]{graphicx}

\begin{document}
\title{Design and cryogenic operation of a hybrid quantum-CMOS circuit}

\author{P. Clapera}
\affiliation{Univ. Grenoble Alpes, INAC-SPSMS, F-38000 Grenoble, France}
\affiliation{CEA, INAC-SPSMS, F-38054 Grenoble, France}
\author{S. Ray}
\affiliation{Univ. Grenoble Alpes, INAC-SPSMS, F-38000 Grenoble, France}
\affiliation{CEA, INAC-SPSMS, F-38054 Grenoble, France}
\author{X. Jehl}
\affiliation{Univ. Grenoble Alpes, INAC-SPSMS, F-38000 Grenoble, France}
\affiliation{CEA, INAC-SPSMS, F-38054 Grenoble, France}
\author{M. Sanquer}
\affiliation{Univ. Grenoble Alpes, INAC-SPSMS, F-38000 Grenoble, France}
\affiliation{CEA, INAC-SPSMS, F-38054 Grenoble, France}
\author{A. Valentian}
\affiliation{Univ. Grenoble Alpes, LETI-DCOS, F-38000 Grenoble, France}
\affiliation{CEA, LETI, Minatec campus, F-38054 Grenoble, France}
\author{S. Barraud}
\affiliation{Univ. Grenoble Alpes, LETI-DCOS, F-38000 Grenoble, France}
\affiliation{CEA, LETI, Minatec campus, F-38054 Grenoble, France}

\begin{abstract}

Silicon-On-Insulator nanowire transistors of very small dimensions exhibit quantum effects like Coulomb blockade or single-dopant transport at low temperature. The same process also yields excellent field-effect transistors (FETs) for larger dimensions, allowing to design integrated circuits. 
Using the same process, we have co-integrated a FET-based ring oscillator circuit operating at cryogenic temperature which generates a radio-frequency (RF) signal on the gate of a nanoscale device showing Coulomb oscillations. We observe rectification of the RF signal, in good agreement with modeling.       
\end{abstract}

\maketitle

While silicon-based complementary metal-oxide-semiconductor (CMOS) technology constitutes the mainstream of electronics, research for the beyond CMOS era focuses mostly on devices relying on new materials and/or quantum features~\cite{itrs2013erd,Koenraad2011,Novoselov2012}. Although circuits can be made with these advanced devices, like graphene-based oscillators ~\cite{Schall2013} or single-electron transistors (SETs)~\cite{Ono2000,Conrad2007,Lee2008a}, a first step is to interface these quantum nanoelectronic devices with conventional CMOS circuits. The integration with mainstream technology is greatly simplified when the novel device is silicon-based. Hybrid circuits using a small number of field effect transistors (FETs) and SETs have been demonstrated~\cite{Inokawa2003,Saitoh2004c,Mahapatra2006,Nishiguchi2006,Zhang2007, Deshpande2012}. Recently an SET device has been integrated with CMOS 1-bit selectors~\cite{Suzuki2013}. In this work SETs working up to 300K were obtained by very small nanowire cross sections resulting in shape fluctuations. Here we demonstrate the integration of an SET relying on well controlled dimensions~\cite{Hofheinz2006a} and a ring oscillator (RO) CMOS circuit designed for low temperature operation, made with more than 600 FETs. The different FET or SET behaviour is obtained by varying the width of transistors all fabricated with the fully-depleted silicon-on-insulator (FD-SOI) nanowire technology~\cite{Barraud2012}. The RO output feeds a non-overlapping clock generators which delivers two, phase shifted square wave signals at radio frequency onto the gates of the SET device. When the RF is turned on we observe a dc current in the SET at zero source-drain bias, due to rectification effect. This effect naturally arises when sufficiently large RF signals are applied to a non-linear device~\cite{Brouwer2001,Giblin2013}.
\begin{figure}
\begin{center}
\includegraphics[width=\columnwidth, clip]{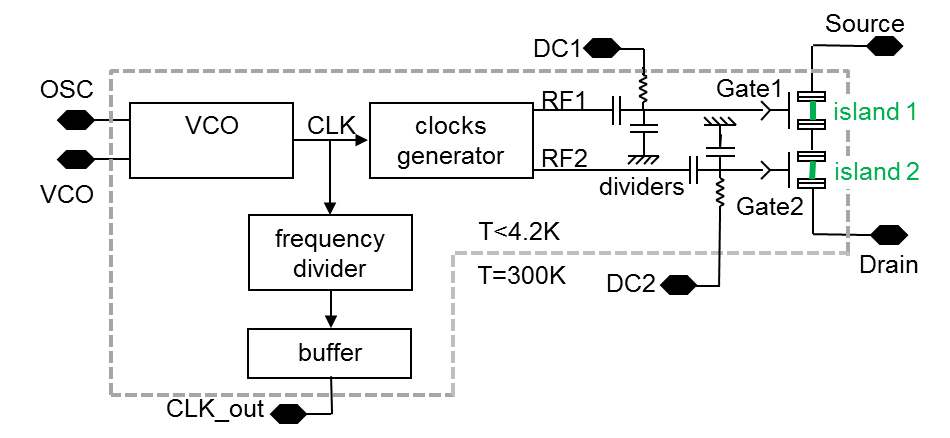}
\caption{Schematics of the whole circuit designed and fabricated on 300\,mm SOI wafers. The nanowire dominated by Coulomb blockade below 10\,K is on the right. The CMOS circuit driving the gates is made of several sub-circuits. A voltage controlled oscillator (VCO), monitored through a frequency divider, feeds a clock generator also based on ring-oscillators. This generator delivers two  delayed and phase-shifted RF signals, $RF1$ and $RF2$, which are further attenuated by capacitive dividers and added to external DC biases. The voltage supply $V_{DD}$ referenced to ground $GND$ is not shown.} 
\label{schematics}
\end{center}
\end{figure} 


A schematic diagram of the whole circuit patterned on a 11\,nm thick SOI film with the trigate SOI technology~\cite{Barraud2012} is shown in Fig.~\ref{schematics}.
The SET is made with a 25\,nm wide nanowire covered by 2 gates of length 40\,nm. Only one of the two gates is used in this study. The gate stack consists of $\approx$ 0.8\,nm of SiO$_2$, $\approx$ 2\,nm HfSiON, 5\,nm of TiN and 50\,nm of polycrystalline silicon. After gate etching a single, self-aligned 12\,nm Si$_3$N$_4$ spacer is deposited. The back-end process follows with epitaxy for raised source/drain, doping and silicidation (NiPtSi). 
\begin{figure*}[!t]
\begin{center}
\includegraphics[scale=0.85, clip]{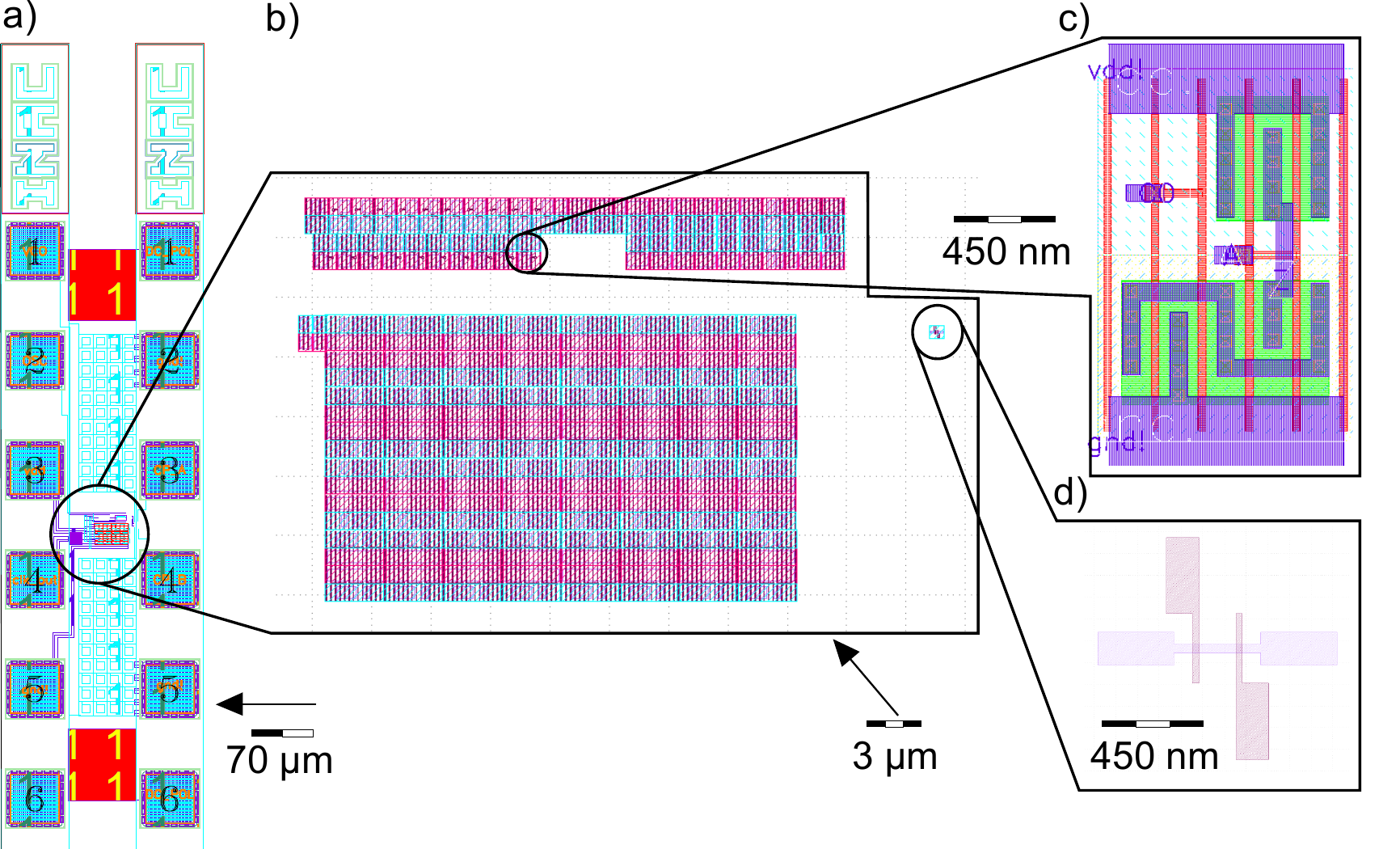}
\caption{\textbf{a)} View of the complete circuit layout with 12 contact pads, the bias resistors in red and capacitors as empty cyan squares between pads 2-3 and 4-5. \textbf{b)} Detailed view of the CMOS circuits located between pads 3 and 4. The top part is the VCO ($\approx$ 130 FETs) and non-overlapping clock generator ($\approx$ 100 FETs), while the bottom part is the frequency divider ($\approx$ 400 FETs). Both are made with 1$\mathrm{\mu}$m wide channels and 60\,nm long gates as depicted in \textbf{c)}. \textbf{c)} FET design with the gate level in red, the source and drain in purple and active area in green. \textbf{d)} Detailed view of the SET device (25\,nm wide channel and 40\,nm long gates).}
\label{mask}
\end{center}
\end{figure*}
The circuit for generating RF signals is made with 2$\times$1\,$\mu$m wide channels and 60\,nm long gates. All the transistors are supplied with a voltage $V_{DD}$ referenced to ground (\textit{GND}). The circuit starts with a Voltage-Controlled Oscillator (VCO) made of 20 inverters and 1 NAND gate. A voltage controlled current source is inserted in the inverters with an N-type FET in footer configuration. The VCO can be switched ON or OFF with the \textit{OSC} input and its frequency is tuned by an external control voltage \textit{VCO}. In agreement with simulations the VCO output frequency ranges from 300\,kHz (\textit{VCO}=0.2\,V) to 1.8\,GHz (\textit{VCO}=1\,V) at 300\,K. It feeds a second, non-overlapping clock generator with two outputs. It is made of 5 buffers allowing to shift by 108\,ps the two signals, ensuring that only one of the two outputs is in the high ($V_{DD}$) state at any time. In addition the frequency of the VCO is monitored by a frequency divider. The two outputs $RF1$ and $RF2$ are attenuated by capacitive dividers in order to lower their amplitudes down to 0.5\,mV, and they are added to two DC voltages $DC1$ and $DC2$ thanks to bias tees realized with 1\,M$\Omega$ poly-silicon resistors. In the end $DC1+RF1$ and $DC2+RF2$ are respectively applied to gates 1 and 2 of the SET device. 

The circuit implementation is shown in Fig.~\ref{mask}, with detailed views of the RO and SET. It uses 12 aluminium pads for external control (see Fig.~\ref{mask}a). The passive components of the circuits are the resistors (red squares) and the 2\,pF and 1\,fF capacitors (empty cyan squares in Fig.~\ref{mask}a) made between the two metal layers of the back-end process. The RO and frequency divider are located between pads 3 and 4 (see Fig.~\ref{mask}a and \ref{mask}b). 

\begin{figure}[!t]
\begin{center}
\includegraphics[width=\columnwidth, clip]{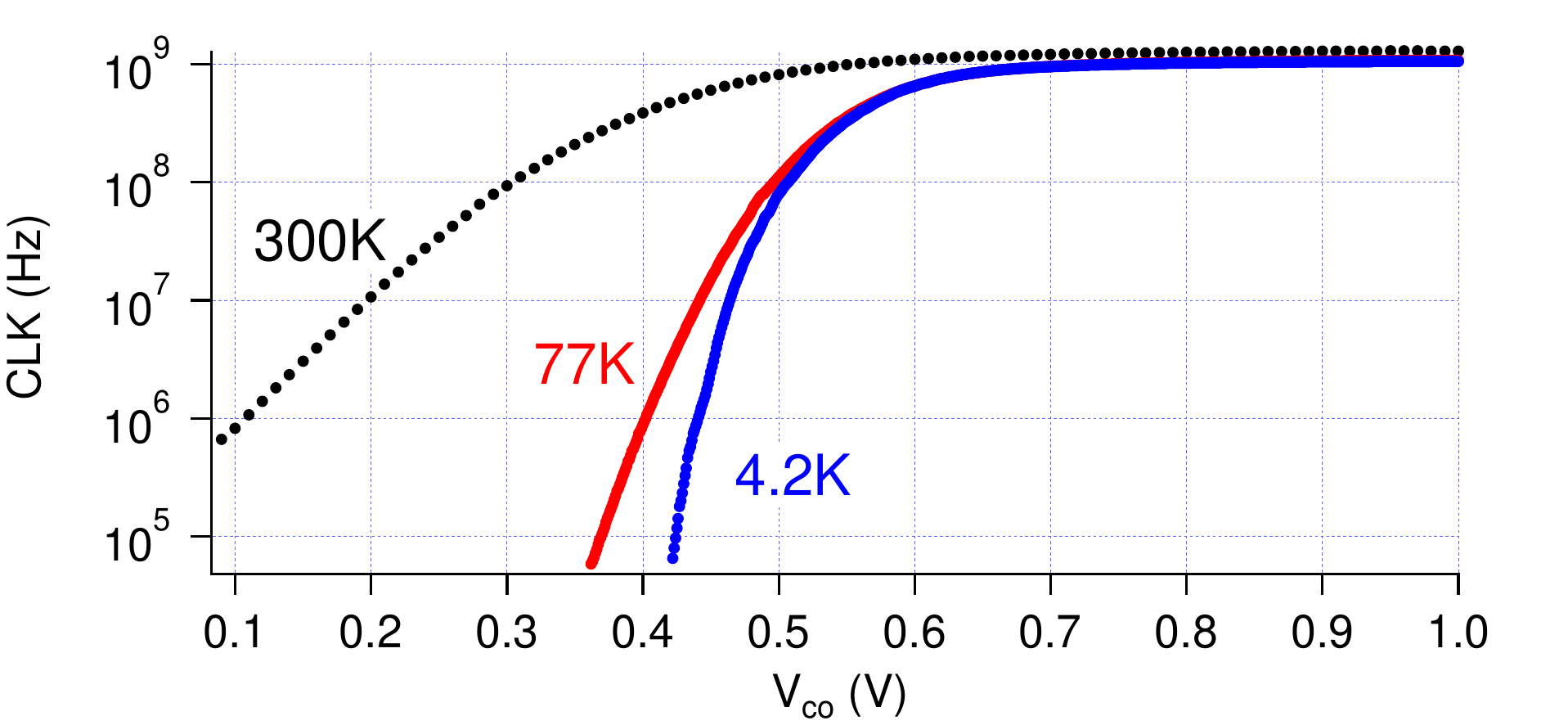}
\caption{Output frequency of the voltage-controlled oscillator measured at 300, 77 and 4.2\,K after correction by the frequency divider (division by a factor of 65536). The maximum output frequency are 1.36\,GHz at 300\,K, 1.08\,GHz at 77\,K and 1.06\,GHz at 4.2\,K.} 
\label{clk}
\end{center}
\end{figure}

The frequency response of the VCO is shown in Fig.~\ref{clk} for various temperatures. The RO being fed by the output current of the N-FETs controlled by \textit{VCO}, we obtain a curve similar to the drain-source vs. gate voltage characteristics of an N-FET, with a sub-threshold regime getting steeper as the temperature decreases and a saturation at 1.36\,GHz at 300K and 1.06\,GHz at 4.2\,K. This very good behaviour down to 4.2\,K of a CMOS circuit containing 600 approximately transistors shows that conventional silicon electronics is perfectly suitable for use at low temperature, provided that passive components such as capacitors and mostly resistors are carefuly designed.

\begin{figure}[!t]
\centering
\includegraphics[width=0.8\linewidth,clip]{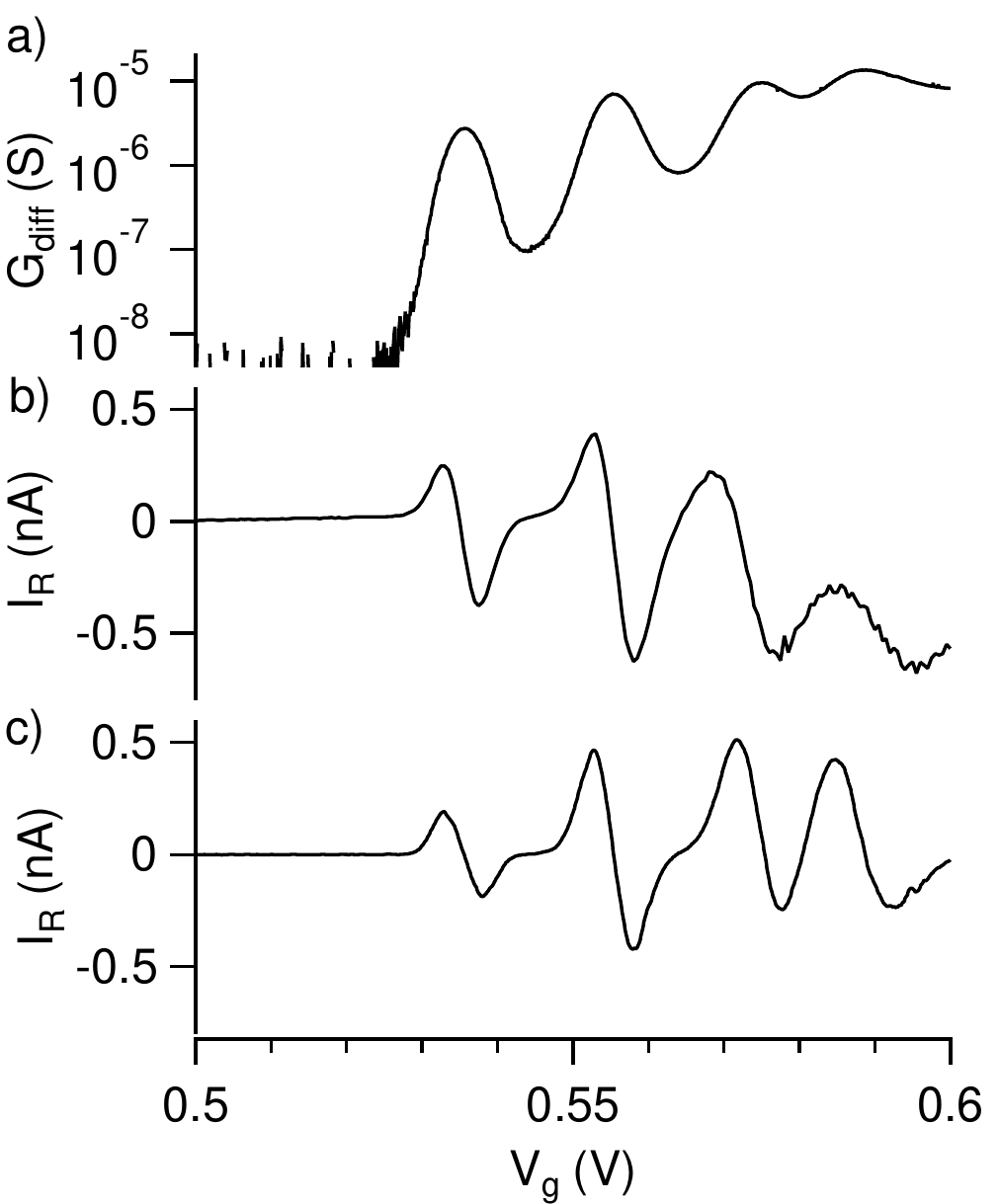}
\caption{\textbf{a)} Coulomb blockade oscillations measured at 1\,K with a lock-in amplifier with an AC signal of 100\,$\mu$V at 77\,Hz. \textbf{b)} DC current measured when $V_{ds}^{DC}$=0 but CMOS circuit switched on, with amplitude 500\,$\mathrm{\mu}$V at 412\,MHz. \textbf{c)} DC current calculated with the rectification model. We find good agreement with the measured current shown in \textbf{b)}. The current follows the derivative of the transconductance, i.e. the second derivative of the conductance.}
\label{data}
\end{figure}

Because of its small cross-section the quantum device driven by the CMOS circuit exhibits Coulomb blockade oscillations below $\approx$ 10K~\cite{Hofheinz2006a, Deshpande2013}, as shown in Fig.~\ref{data}a where the low-frequency transconductance $G_{diff}$ versus the gate voltage which is varied ($V_{g}$) is shown at 1\,K. Four quasi-periodic peaks are observed, corresponding to the addition of 4 electrons in the channel below the gate. The period of 18\,mV in $V_{g}$ corresponds to an effective gate capacitance of 9\,aF. These results are obtained without DC drain-source bias and no RF applied, hence there is no DC current flowing through the device in that case.

When the CMOS circuit is turned on, but still no DC bias applied, we measure a DC current, shown in Fig~\ref{data}b. The presence of current in absence of bias and its characteristic dependence with $V_g$ is well explained by a rectification effect. In nanoscale devices which are by principle difficult to contact perfectly, AC signals driven onto gates can induce a parasitic oscillatory source-drain bias~\cite{Brouwer2001}. For electron pumping experiments it is important to discriminate between this spurious current and the true pumped current~\cite{Brouwer2001,Giblin2013}. Following these previous studies, we consider an RF driven gate voltage $V_g(t)=V_{g}^{DC}+Asin(2\pi ft)$ which couples capacitively to the source and drain, hence creating an additional AC bias component at the same frequency $f$ in addition to the DC bias $V_{ds}^{DC}$: 
\begin{equation}
V_{ds}(t)=V_{ds}^{DC}+kV_{ds}^{AC} sin(2\pi ft+\phi),
\label{Vds}
\end{equation}
where $k$ and $\phi$ characterize the coupling.
The rectified current is the average over one period $\frac{1}{f}$ of the resulting current $I(t)=V_{ds}(t)G(t)$:
\begin{equation}
I_R= f \int_{0}^{1/f} V_{ds}(t)G(t)dt
\label{Ir}
\end{equation}

As already pointed out in~\cite{Brouwer2001} and~\cite{Giblin2013}, the general expression obtained by combining \ref{Vds} and \ref{Ir} is greatly simplified in the limit of small driving amplitude $A$. This is the standard case for AC lock-in measurements where one can use a linear approximation: $I(t)\propto \frac{\partial G}{\partial V_g}|_{V_g=V_g^{DC}}$. Here we are interested in the integral over one period (equation~\ref{Ir}) to get the DC component, hence the same approximation is used again and  
\begin{equation}
I_R\propto \frac{\partial ^2 G}{\partial V_g^2}|_{V_g=V_g^{DC}}.
\label{derivative}
\end{equation}

This model is used to calculate the rectified current $I_R$ in the general case and taking into account our non-sinusoidal RF excitation by using the Fourier series describing a square wave instead of a single sine wave. The results are shown in Fig~\ref{data}c. We found an excellent agreement with the measurements (Fig~\ref{data}b) and recover the result that $I_R$ is proportional to the second derivative of the conductance (equation \ref{derivative}). This is expected since we operate the circuit with RF output amplitude 500\,$\mathrm \mu$V, which is small compared to the Coulomb oscillations period in the transconductance. Indeed this amplitude is of the order of the linewidth of the graph in Fig~\ref{data}a.

\section{Conclusion}

We have designed, fabricated and operated down to 4.2K a circuit allowing to generate on-chip RF signals on the gates of a nanoscale quantum device. The clock generators based on a ring oscillator as well as the capacitance divider and bias resistors are fully operational down to 1\,K. We have observed a finite DC current through the quantum device in the absence of DC bias when the RF drive is turned on. This current which scales with the derivative of the differential conductance is well understood within the framework of rectification due to capacitive coupling of the gate signal to the source and drain of the nanodevice. These results pave the way for the integration of conventional CMOS circuits operating at low temperatures together with quantum devices.


P. Clapera acknowledges support from the PhD program of the Nanosciences foundation in Grenoble. This work was partially supported by the EU through the FP7 ICT projects TOLOP (318397) and SiAM (610637), and by the Joint Research Project ”Qu-Ampere” (SIB07) from the European Metrology Research Programme (EMRP). The EMRP is jointly funded by the EMRP participating countries within EURAMET and the European Union.

\end{document}